# A hybrid econometric-machine learning approach for relative importance analysis: Prioritizing food policy

**Akash Malhotra** [a,1]

[a] *Centre for Economic Studies and Planning, Jawaharlal Nehru University, New Delhi, India*

## Abstract

*A measure of relative importance of variables is often desired by researchers when the explanatory aspects of econometric methods are of interest. To this end, the author briefly reviews the limitations of conventional econometrics in constructing a reliable measure of variable importance. The author highlights the relative stature of explanatory and predictive analysis in economics and the emergence of fruitful collaborations between econometrics and computer science. Learning lessons from both, the author proposes a hybrid approach based on conventional econometrics and advanced machine learning (ML) algorithms, which are otherwise, used in predictive analytics. The purpose of this article is two-fold: to propose a hybrid approach to assess relative importance and demonstrate its applicability in addressing policy priority issues with an example of food inflation in India, followed by a broader aim to introduce the possibility of conflation of ML and conventional econometrics to an audience of researchers in economics and social sciences, in general.*

[1] *Corresponding author at:* Centre for Economic Studies and Planning, Jawaharlal Nehru University, New Delhi-110076, India.

*E-mail address:* akash_malhotra@iitb.ac.in

# 1. Introduction

One of the most common requests made by economists to statisticians is for a measure of relative importance[2] of independent variables in their econometric models. For example, health economists are interested in assessing the relative importance of various socioeconomic variables like income, wealth, social class, education etc. in determining health inequalities (Mondal and Shitan, 2014, Kjøllesdal et al., 2010). Although there are no unambiguous measures of relative importance, standardized regression coefficients and zero-order correlations are often used to answer this question by researchers, but they have been shown to fail in considering both the effect the variable has by itself and in presence of other variables in the model (Johnson, 2000).

There are two approaches in statistics for making inferences from data: one assumes that data are generated from a given stochastic process and then builds the statistical model, while the other focusses on algorithmic models and treats the data generating mechanisms as unknown. Econometricians tend to subscribe to the former and it is only recently that economists have started exploring algorithmic approaches. In this backdrop, I propose a hybrid approach based on a conflation of Machine Learning (ML) and conventional econometrics to assess the relative importance of independent variables. Here, I demonstrate applicability of the proposed hybrid econometric-ML approach to already-established determinants of food inflation in India, where statistically significant independent variables are first identified through common econometric techniques and those variables are then used in constructing an exploratory (no independent testing) model using machine learning techniques which provides an opportunity to quantify the relative variable importance.

[2] It is important to realize the difference between variable selection and the exercise of ascribing relative importance to independent variables. Variable selection (or feature selection) is a pre-processing exercise for which a long list of references of regularized regression models exists in the literature. Determining relative importance on the other hand is a post-hoc exercise, the literature for which is still underdeveloped.

Unlike other domains of science, the adoption of ML in economic research has been sparse (Einav and Levin, 2014) and slow, but now these techniques are being adopted in empirical work and are the topic of an expanding methodological literature (Athey & Imbens, 2019). ML techniques could potentially serve as a powerful econometric tool in estimating exploratory/predictive economic models on high-dimensional data (Varian, 2014). In the future, ML is expected to become a standard tool of empirical research in economics as well as contribute to the development of economic theory. Development of new methods combining conventional econometrics and ML for solving traditional estimation problems in social science has begun (Athey & Imbens, 2019; Storm et al., 2020) and I predict that this literature will expand remarkably over the next two decades.

The rest of the article is organized as follows: Section 2 presents the limitations of existing approaches in determining the relative importance of variables. Section 3 expounds the hybrid approach central to this article. Section 4 illustrates the applicability of the proposed approach in assessing the relative importance of determinants of food inflation with appended takeaways for food policy in the Indian context. Section 5 concludes with caveats and suggestions for future research.

# 2. Relative Importance: Limitations of conventional econometrics

Economists often have the dual goal of prediction and explanation which are followed by an attempt to understand the importance/significance of predictor/independent variables. The search for a method to rank and quantify the relative importance of independent variables is, however, going on for many years (Johnson & LeBreton, 2004). The statistical method most commonly employed by economists are regressions (Ramcharan, 2006). However, the commonly reported statistics within regression results do not answer the question of the relative importance of independent variables. Regular regression coefficients denote the mean change in the dependent variable given a unit increase in the independent variable. However, larger coefficients don't necessarily imply more variable importance as the units of measurement may vary between the different type of variables. Another commonly reported statistic, *p-value*, which is used to establish *statistical significance* is unusable for determining variable importance as low *p-values* could also arise from large sample sizes or very precise estimates and hence, don't necessarily indicate relative importance.

To overcome these limitations, researchers in the past have developed straightforward methods for ranking variables based on their relative importance, with standardized regression coefficients (or *beta weights*) being the most commonly used (Darlington, 1990). Beta weights represent the mean change in the dependent variable given a one standard deviation change in the independent variable, allowing them to be compared across different type of variables. The sum-product of independent variables' beta weights and zero-order correlation with dependent variable is equal to $R^2$ (R-squared) of the model. When predictors are uncorrelated, beta weights and zero-order correlations are equivalent so, in such a case the relative importance of each variable can be evaluated from beta weights. *Usefulness* is another measure commonly used to ascribe relative importance to variables based on the increase in $R^2$ that a variable produces when it is added to a model that already contains all the other variables (Darlington, 1968).

Unfortunately, these simple measures for analyzing the relative importance of independent variables fail to properly partition variance to different variables when the variables are correlated with one other, which is often the case with economic time series data. In the presence of multicollinearity, relative contributions of the independent variables having the weakest correlation with dependent variable may be ignored or subdued in the model, which can even lead to negative regression coefficients for variables that are indeed positively correlated with independent variable (Stadler et al., 2017). A rich literature has developed on assessing the relative importance of independent variables in the fields of psychology, economics, statistics, medicine and consumer science, which has been critically reviewed by Johnson and LeBreton, (2004). The two approaches, *dominance analysis* (Azen & Budescu, 2003) and *relative weights analysis* (Johnson, 2000), have become widely popular in the recent past. Both these methods take into account an independent variable's direct effect and its effect on outcome in presence of other variables, and produce relative importance scores that signify the proportionate contribution each variable makes to $R^2$.

Dominance analysis requires specification of all possible subsets of a regression model and establishment of general dominance of variables; the average increase in $R^2$ associated with a variable across all possible submodels quantifies variable importance. Dominance analysis approach tends to become cumbersome and computationally expensive as the number of independent variables increases; a model with (n) independent variables require specification of ($2^n$ – 1) submodels. Johnson (2000) therefore suggested relative weights analysis as an alternative wherein correlated independent variables are transformed into *maximally related orthogonal variables*, i.e., transformed variables that are uncorrelated with each other but maximally correlated with the original dependent variable. It is important to understand that relative weights approach does not solve the problem of multicollinearity, it just circumvents it, and acknowledging that, Johnson (2000) itself points out a major shortcoming of this approach with highly correlated independent variables — when multicollinear variables have similar correlations with dependent variable also, relative weights could get artificially inflated. In this regard, Stadler et al., 2017 recommends dropping one of the highly multicollinear variables but unfortunately, no absolute cut-off exists to distinguish too much multicollinearity (see Craney & Surles, 2002 for a discussion of cut-off values). Moreover, this approach is not invariant to the choice of orthogonalizing procedure; it is possible to orthogonalize in infinitely many ways and Johnson (2000) defends his choice of least squares orthogonalization by comparing the relative importance results (in terms of similarity) with those obtained by dominance analysis — which is not much of a compelling argument. Johnson (2000) points out that these relative weights are likely to change if some of the independent variables are replaced with a linear combination of those variables. Finally, comparing two relative weights, like any other statistic, requires estimation of their respective confidence intervals as the magnitude of correlation depends to some extent on the reliability of the measurement.

Speaking of hybrid approaches, Belloni et al. (2014a) develop a novel estimation and inference method for the treatment effect in a high-dimensional setting, called the "post-double-selection" method. The method begins with employing ML algorithms for selecting a set of control variables that are useful for predicting the treatment, followed by a second step of selecting additional variables that predict the outcome variable. In the final step, the authors estimate the treatment effect of interest by regressing the outcome variable on the treatment and the union of the set of variables selected in the previous two variable selection steps. Using the aforementioned "post-double-selection" method with LASSO, Dutt & Tsetlin (2020) examine the relative importance of inequality vs poverty for development outcomes in a causal inference framework and find poverty headcount measure to be more important than inequality (as measured by Gini). Although the approach has been named as "post-double-selection" method, the reader may have noticed from the sequencing of steps outlined in the procedure by Belloni et al. (2014a) and the usage of LASSO by Dutt & Tsetlin (2020) that this approach is more close to the pre-processing exercise of variable selection than the post-hoc exercise of ascribing relative importance. Moreover, the approach is incapable of quantifying relative importance through normalized scoring or any other procedure.

# 3. A hybrid approach

Generally, time series data in economics is available at lower frequencies (quarterly or annually) and the farther we go back in time the less likely it becomes to find data of reliable quality. Consequently, this dearth of macroeconomic data makes predictive modeling extremely challenging in economics which usually performs well on 'big data'. Hence, the optimal approach[3], which I recommend here, is to use established econometric techniques for drawing causal inferences and ascertaining statistical significance of independent variables using typically ‘small’ economic datasets and then build on those results with machine learning algorithms which are commonly employed in predictive analytics.

[3] In this paper, I have attempted to adhere to recommendations made by Athey & Imbens (2019) — while using ML tools, researchers should clearly articulate the goals of their analysis and accordingly

Now, I propose a straightforward 3-step hybrid approach to quantify the relative importance of variables:

**Step 1**: *Identify statistically significant variables which explain the dependent variable from common econometric techniques (generally, multiple regression for cross-sectional data and Vector Autoregressive models for time series data).*

**Step 2**: *Construct an exploratory (no independent testing) model with variables identified in step 1 using machine learning algorithms which are commonly employed in predictive analytics.*

**Step 3**: *Now, to quantify the relative importance of independent variables, employ specific feature importance modules of machine learning algorithms for data mining in the exploratory model constructed in step 2.*

As a side note, it’s important to clarify the distinctions between “*predictive analytics*”, “*predictive modelling*”, “*data mining*”, “*machine learning*” and *“Artificial Intelligence”* which are commonly misconstrued as synonyms. *Artificial Intelligence* (AI) focusses on designing systems that could imitate human decision-making processes and carry out tasks in ever more human ways. *Machine learning* encompasses various algorithms which can recognize patterns in data and learn from them; and serves as a tool to teach computer systems to learn for themselves and thence become “artificially intelligent”. *Predictive analytics* includes a variety of statistical techniques, viz., data mining, predictive modeling and machine learning. *Predictive modeling* is the use of conventional statistics or machine learning algorithms to

---

clarify why certain properties of the ML algorithm in use and the corresponding results may or may not be important for the objectives of the analysis.

estimate future outcomes. And finally, *data mining*, which is more of a cross-disciplinary field, uses conventional statistical techniques, visualization, as well as, machine learning algorithms to discover and extract properties from datasets.

The proposal of the above hybrid approach has been motivated by an important observation made by Shmueli (2010) that the statistical models used for causal explanation in social sciences are almost always association-based models[4], of which regression is the most common example. Although the concept of causality has shown to be evasive in the context of social sciences when trying to formalize it, there exists mainly two approaches of practical relevance:

i. *Physical Influence:* This approach is used for making causal inference over random variables and is commonly employed in the case of cross-sectional data. It is usually denoted as an ***A/B* test** wherein the manipulation of the cause changes the effect.

ii. *Temporal Precedence:* This approach is used for making causal inference over stochastic processes and is commonly employed in the case of time series data. It is based on comparing the ability to predict *Y* from all the information in the Universe, denoted as *U*, with the prediction of *Y* obtained from all the information in *U* except for some stochastic process *X*. The approach is usually denoted with ***U/X***.

The critical step in establishing causality within the *A/B* framework is estimating the counterfactual — a prediction of what would have happened in the absence of an intervention. In this regard, Varian (2016) proposes an elegant approach as an alternative to the commonly used control group data as an estimate of the counterfactual; the proposed train-test-treat-compare (TTTC) method employs standard ML tools to build a predictive model using data from before the experiment was run and the counterfactual is provided by the predictions of this model, thus eliminating the need for control groups. For time series analysis, the causal inference is mostly done within the *U/X* framework and the interpretation of causality proposed by Granger (1969, 1980) is most widely used concept, which posits that the effect is preceded by its cause (Eichler, 2012). It is important to recognize the subtle differences between the theoretical notion of Granger-causality as outlined in Granger (1969, 1980) and its empirical versions commonly referred to as "Granger-causality". The core idea behind Granger causality is that the causal time series, *X,* contains some unique information about the series being caused, *Y*, that is not manifested in the Universe which excludes *X*, $U - X$. Practically, if removing *X* from the model reduces it predictive power regarding *Y*, then we infer that *X* "Granger-causes" *Y*. It is important for the reader to know that Granger-causality as inferred from common econometric techniques is primarily a measure of association between variables and could even lead to spurious causalities if important variables are omitted from the model (Hsiao 1982).

[4] To the dislike of statisticians, social scientists prefer association-based models for explanatory modelling although “proper” statistical methodology, such as randomized experiments or specialized causal inference methods for non-experimental data, for testing causality do exist (Shmueli, 2010).

In economics (and many other scientific fields such as psychology, education, etc.), it is posited that models that possess high explanatory power often command inherent predictive power. The reason behind this presumption is many-fold, Firstly, econometrics in its early stages of development (early and mid-20$^{th}$ century) seem to be heavily influenced by the then-contemporaneous philosophy of science literature, particularly the hypothetico-deductive method of Hempel & Oppenheim (1948), which explicitly equates prediction and explanation. Secondly, the role of theory in economics is central to its development and consequently the discipline views data and statistical modeling strictly through the lens of the theoretical model and the associated restrictions — demand monotonicity, discrete choice among substitutable products, to name but a few. But what comes as a surprise is economists' reluctance to go the other way, i.e., employ predictive modeling for theory building or testing and then ascribe causation. This is quite absurd given the fact that explanatory models can never be confirmed and are difficult to contradict due to the absence of an unbiased estimate of the ground truth, whereas predictive models could be accepted/rejected based on negative empiricism. A possible reason for this bias against predictive analytics could be the statistical training of non-statistician researchers which is reflected in the standard introductory statistics textbooks (Shmueli, 2010). A deeper cause of this bias lies in the social science roots of economics as a discipline, where prediction is largely considered unscientific (Berk, 2016). Economists keep on disagreeing among themselves whether prediction per se is a legitimate objective of economic science (Feelders, 2002; Athey & Imbens 2019) and, are even skeptical about these newer data mining methods which put more emphasis on predictive fit (Einav and Levin, 2014). This observed resistance to adoption of ML methods comes from the extant academic culture: Economics journals emphasize the use of conventional econometric methods that come with formal properties like large sample properties of estimators and tests, which include consistency, normality, and efficiency. A majority of the ML methods do not naturally deliver the type of properties desired by economics journals and instead, focus on working properties of algorithms and the performance of the algorithm in a given setting, e.g., error rates. Moreover, the ability to construct sample confidence intervals and calculate p-values is considered highly important in the economic literature; the editors of economics journals should reconsider this requirement and refrain from outrightly brushing-off methods that cannot deliver such post-hoc analyses (or, possibly, cannot deliver them yet, as is the case with ML) if these methods offer other advantages and insights.

Over the last few years, the discipline of economics has witnessed considerable advancements related to the use of ML. Athey & Imbens (2019) provide an excellent review of specific methods from ML literature as well as newly developed methods at the intersection of ML and econometrics, which, in the view of authors, are important to know for econometricians and empirical economists. In the context of agricultural economics and food policy, Storm et al. (2020) provide a useful overview of ML approaches and demonstrate that ML is in many ways a natural extension of econometrics toolbox of an agricultural economist; the current set of econometric tools fail to offer the flexibility needed to handle the multiple interactions, non-linearities and heterogeneities endemic to environmental or social processes which characterize the agricultural systems. The bulk of literature has focussed on causal inference. Athey (2019) presents an overview of the themes in the emerging econometric literature combining machine learning and causal inference and describes the implications of ML for the discipline of economics, as a whole. Belloni et al. (2014b) provide an overview of how methods or algorithms used in predictive modelling can be adapted and modified to provide inference about model parameters in the context of high-dimensional data. Chernozhukov et al. (2018) propose an approach which they named double or debiased ML (DML). The approach employs ML methods in the context of the classic semi-parametric problem of inference on a low-dimensional parameter (typically a causal parameter or treatment effect parameter) in the presence of a high-dimensional nuisance parameter. Wager & Athey (2018) extend Breiman's widely used random forest algorithm and develop a non-parametric 'causal forest' for estimating the heterogeneity in treatment effect.

By snubbing predictive analytics, economics, as a discipline, might have been missing out on new causal mechanisms and new hypotheses, which data mining techniques and advanced machine learning algorithms can unearth by capturing underlying complex patterns and relationships. An exception to this trend is manifested in Kleinberg et al. (2015), wherein authors explicitly argue in favour of ML over traditional econometric approaches in solving prediction problems and illustrate the benefits of solving prediction policy problems in the form of not only enhanced social impact but also richer theoretical and economic insights. To fill this lacuna and encourage conflation of explanatory and predictive analytics in economics, this article has proposed a simultaneous use of data mining techniques usually employed in predictive analytics and commonly used explanatory models in econometrics. This marriage between explanatory econometrics and predictive analytics is especially convenient in the case of relative importance assessment in association-based regression models, as, in predictive analytics, the predictors are selected based on their ex-ante availability and quality of association between predictors (analogous to independent variables) and the response variable (analogous to dependent variable). In such a case, using machine algorithms for assessing the relative importance of variables whose statistical significance has already been established by conventional econometrics should prove to be efficacious.

# 4. Illustrative Example: Food Inflation in India

This section demonstrates the application and utility of the hybrid approach with food inflation in India as the target variable. The following independent variables were chosen for *step 2*, after establishing their statistical significance in explaining food inflation in India as part of the *step 1*: Minimum Support Prices (MSP), international food prices, fiscal policy, farm wages, agricultural input prices, performance of southwest monsoon, and share of protein-rich food items. The reader may note that the statistical significance of the aforementioned variables has been repeatedly established in the literature (see Table 1).

**Table 1** Econometric studies on food inflation establishing statistical significance of regressors

| | MSP | FWI | AgriInput | FAO | ProteinExp | FD | MonsDev |
|---|---|---|---|---|---|---|---|
| Author's estimate (VAR-based tests for Granger causality) | ✓ | ✓ | ✓ | ✓ | ✓ | ✓ | ✓ |
| Sonna et al. (2014) | ✓ | ✓ | ✓ | – | ✓ | – | – |
| Gulati & Saini (2013) | – | ✓ | – | ✓ | – | ✓ | – |
| Mohanty (2010) | – | – | – | – | – | – | ✓ |
| Varma & Saraf (2016) | ✓ | ✓ | ✓ | ✓ | – | – | × |

When dealing with time series data, Granger-causality has been the preferred vehicle for establishing causal inference in the literature (Eichler, 2012). Granger-causality, by virtue of its definition, offers ease in formulation and flexibility in testing as it could be tested in both parametric and non-parametric time series models. The conventional framework to test Granger-causality has been Vector Autoregressive (VAR) models (Eichler, 2006; 2012). Therefore, VAR-based test for Granger-causality was performed on the given dataset as part of the *step 1*. The general outline of the procedure is as follows: Ensure stationarity of all the series (if not, stationarize them through detrending or/and differencing). Specify a restricted model in which $Y$ is assumed to be dependent only on the lagged values of itself with linear coefficients $\gamma_i$ and a time-dependent noise term $e_t$:

$$Y_t = \gamma_0 + \sum_{i=1}^{p} \gamma_i Y_{t-i} + e_t$$

Now, specify an unrestricted model in which $Y$ is assumed to be dependent on lagged values of $Y$ and $X$, determined by coefficients $\alpha_i$, $\beta_i$ and a time dependent noise term $u_t$.

$$Y_t = \alpha_0 + \sum_{i=1}^{p} \alpha_i Y_{t-i} + \sum_{i=1}^{p} \beta_i X_{t-i} + u_t$$

Finally, the null hypothesis ($H_0: \beta_i = 0, \forall\ i$) is tested. This is to statistically test if $X$ add unique information regarding $Y$ or the unrestricted model provide a better description of $Y$ when compared to the restricted model of $Y$. If the null hypothesis is rejected, we say that $X$ Granger-causes $Y$.

As an illustration of the proposed hybrid approach, a nonparametric regression ensemble has been developed in *step 2* using decision trees with gradient boosting for least squares (Friedman 2001) to assess the relative importance of individual features (analogous to independent variables/predictors) in explaining the inflationary trend in food prices. Unlike parametric regression, supervised machine learning techniques, such as gradient boosting, do not attempt to characterize the relationship between predictors and response with model parameters but rather produces an ensemble of weak prediction models, decision trees in this case, such that it improves the overall performance of model (Hastie et al., 2009). Boosting, as a concept, provides sequential learning of the predictors[5], where each decision tree is dependent on the performance of prior trees and learns by fitting the residuals of preceding trees. The final BRT model is a linear combination of multiple trees that could be thought of as a regression model where each term is a tree (Elith et al., 2008). Unlike linear regression, regression with boosted decision trees can model complex functions by accounting for nonlinearity and interactions between predictor variables without the need of specification of the form of that non-linearity beforehand (Müller et al. 2013). While a single model may be susceptible to noise, ensemble learning minimizes its impact (Sabzevari 2015) — remember, this was one of the main limitations of relative weights approach. The ability of BRTs to automatically handle missing data points saves the effort of data pre-processing. The mathematical formulation of BRTs and the parameters involved in its calibration are described in *Appendix A*.

[5] Readers must refrain from interpreting the terms, *"predictors"* and *"prediction models"*, in their conventional machine learning based predictive modelling sense, where the aim is to predict future outcomes, and should instead, construe these terms as independent variables and association-based exploratory models respectively. However, I continue to use the former terminology for the sole purpose of maintaining consistency with the literature referred to in this section of the article.

In machine learning, there happen to be two types of approaches to assess relative importance of features (*step 3*), one, model-specific and the other being, model-agnostic. In the context of tree-based ensemble methods such as BRTs, a naïve way to measure variable importance is to look at *'selection frequency'*, i.e., the number of times each variable is selected by all individual trees in the ensemble. However, in this article I employ a more elaborate approach (as proposed in Friedman & Meulman, 2003) wherein the relative importance of the predictor variables in the model is assessed via a relative influence score of a variable, which is decided based on the number of times a variable gets selected during splitting, weighted by the squared improvements, and averaged over all possible trees. The cumulative sum of improvements of all splits associated with the given variable across all trees gives the raw variable importance. These raw scores are then used to rescale the variable importance values such that the most important variable gets a relative score of 100, while the other variables are rescaled to reflect their importance relative to the most important variable on a scale of 0 to 100. For a detailed explanation, the reader may refer to Friedman & Meulman (2003).

A different kind of approach, called *'permutation importance'*, is prevalent in machine learning literature and this approach, being model-agnostic, could even be used for so-called "black-box" models like neural networks, where it is difficult to explain how the model characterizes the relationship between the features and the response variable. Inspired by seminal paper on random forests by Breiman (2001), the rationale behind this measure is pretty straightforward: When the independent variable $X_i$ is randomly permuted, its original association with the response Y get broken. If the original variable $X_i$ was associated with the response Y, then altering or permuting its values should result in a substantial decrease in the model performance. The importance score quantifies the contribution of variable $X_i$ to the performance of the model by measuring the reduction in performance after permuting the feature values while keeping the remaining independent variables unperturbed based on a pre-specified evaluation metric. When the evaluation metric used is an error/loss metric (such as Relative squared error, Root mean squared error etc.), then the permutation importance score ($Pis$) is defined as,

$$Pis = -(P_s - P_b)$$

where $P_b$ is the base performance metric score and $P_s$ is the performance metric score after shuffling. This is done iteratively for each of the independent variables, one at a time. The permutation scores are then scaled on 0 to 100 in a fashion similar to splitting-based importance score discussed above.

## 4.1 Data

In the present example, the annual inflation in retail food prices (FCPI), measured by year-on-year change in prices of food items included in CPI-IW (base 2001), is chosen as the response variable, whereas the independent variables include:

| | |
|---|---|
| MonsDev | : Deviation in rainfall received during southwest monsoon in a year from its 50-year long-term mean |
| MSP | : YoY change in production weighted MSP index of major food crops |
| FAO | : YoY change in FAO price index denominated in INR |
| FD | : Combined fiscal deficit as a percentage of total GDP in a year |
| FWI | : YoY change in farm wage index |

AgriInput : YoY change in price index of agricultural inputs constructed from WPI

ProteinExp : YoY change in ratio of private consumption expenditure on protein-rich food items to total food expenditure

The annual data series is considered for all the variables spanning a period of 25 years starting from FY91 to FY16. Data used in this study have been primarily imported from various official sources, viz, Reserve Bank of India (RBI), Labour Bureau of India, Central Statistics Office (CSO), Ministry of Agriculture & Farmers Welfare, Ministry of Finance and Open Government Data (OGD) Platform India. A production weighted MSP index (base: FY05) has been created for major food crops including; cereals - rice, wheat, and maize; pulses - gram and tur; oilseeds - groundnut, mustard, and soybean. The annual average of global food price index (base: 2002-04) published monthly by FAO (Food and Agricultural Organisation) is converted into rupee terms using the yearly average of USD-INR exchange rate. Combined fiscal deficit of central and state governments as a percentage of total GDP is considered. A farm wage index[6] (base: FY05) has been constructed by aggregating wages earned by agricultural laborers engaged in primary farm operations, namely, ploughing, sowing, transplanting, weeding, and harvesting. The weights used in constructing non-labour agri-input price index[7] (base: FY05) have been derived from WPI 2005 series. Relative expenditure on protein-rich items is based upon weights derived from Private Final Consumption Expenditure[8] (PFCE) at constant prices (base: 2003-05), for both plant and animal protein-based items, namely, pulses, oil & oilseeds, milk & milk products, and meat-egg-fish (MFE). Before moving on to results from the proposed hybrid approach, it would be prudent to familiarize ourselves with the background of food inflation in India to better understand the practical relevance of the variables under study here.

[6] The data for farm wages is available from FY91, hence YoY change could only be measured starting from FY92.

[7] The non-labour agricultural inputs include agricultural electricity, light diesel oil, high speed diesel, agricultural machinery & implements, tractors, lubricants, fertilisers, pesticides, cattle feed and fodder.

[8] PFCE item-wise breakup data is available only till FY13.

## 4.2 Country context and background

Unlike developed economies, food inflation[9] has had a significant impact on cumulative inflation, as food expenditure constitutes more than 40 percent[10] of total household expenditure in India. Consequently, future inflation expectations are driven, to a great degree, by food prices in this country - creating a vicious circle. The retail food inflation has grown at an average rate of 9.82 percent between FY07 – FY 16 and even crossed double digits in four instances (Fig. 1), with food prices becoming more than double in absolute terms in these last ten years.

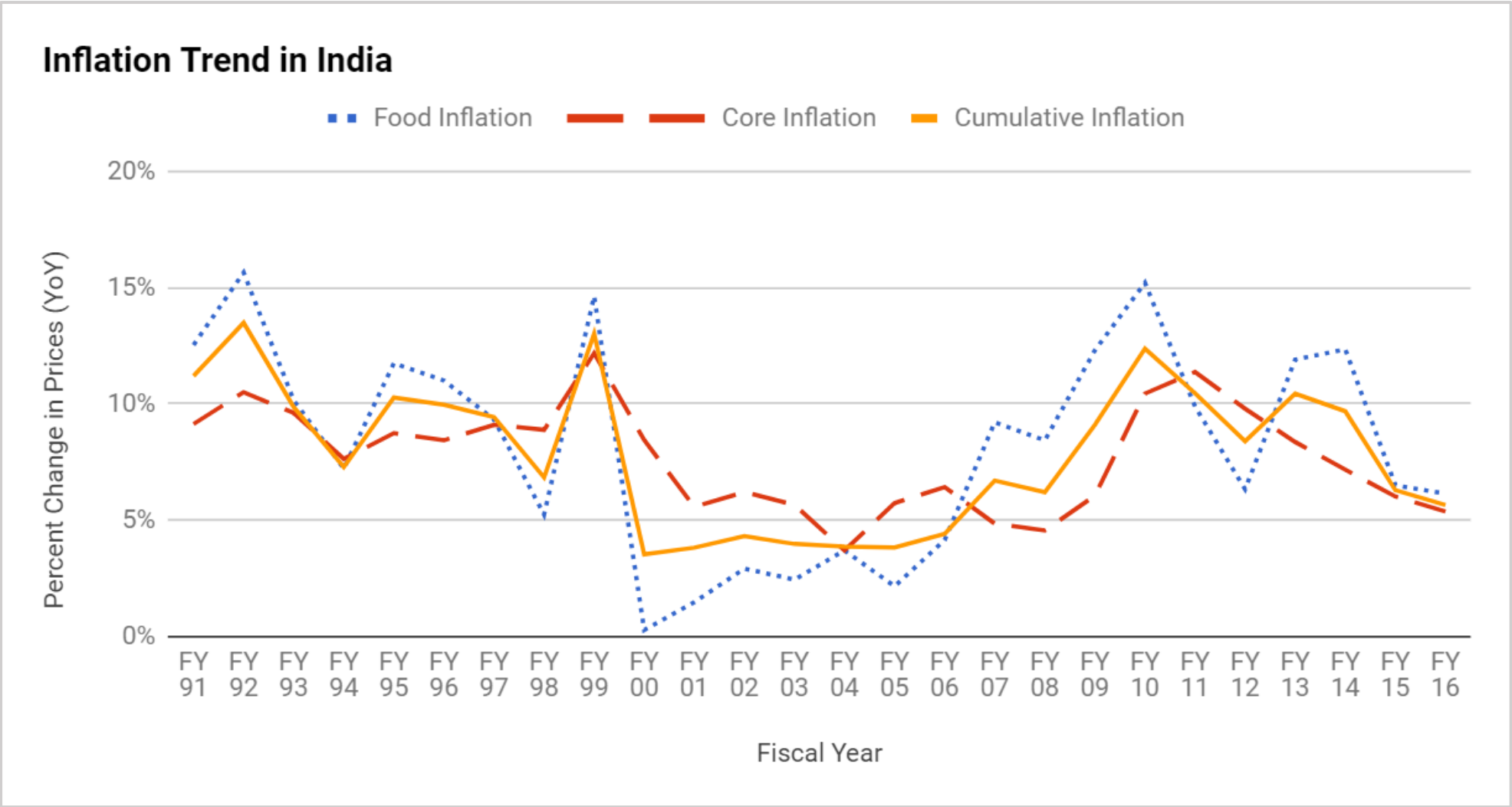

**Figure 1.** Inflation trend based on CPI-IW during FY91-FY16. *Source*: DBIE, RBI

After episodes of food crisis in 1970s, India has followed the path of achieving self-sustenance in two main staples - rice and wheat, complemented by centralised procurement of these two crops from the market to meet the needs of buffer stocks and grain distribution system run by central and state governments which delivers rice and wheat to poor consumers at highly subsidised prices. The government exercises control over this policy through twin instruments, viz, Minimum Support Prices (*MSP)* for cultivators and Public Distribution System (*PDS*) for economically weaker sections. The current structure of Food Administration in India has been neatly summarised by Saini and Kozicka (2014, pp. 9-14). The MSP for eligible crops is decided and announced at the beginning of sowing season by central government; the MSP as a policy instrument is designed to be the national floor level price at which Food Corporation of India (FCI) procures or buy whatever quantities farmers have to offer. India cautiously regulates its agricultural trade by frequently imposing import-export bans and change in duties to protect the interests of domestic farmers and relevance of MSPs as floor prices and, of course, to shield the domestic markets from global price bouts. In the wake of 2007-08 global food price crisis, India adopted a very restrictive trade policy including a ban on exports of wheat and rice which continued till 2011.

[9] For this study, retail inflation is based on CPI-IW for industrial workers, unless stated otherwise.
[10] *Source:* 68th NSSO (National Sample Survey Office) Consumption Expenditure Survey 2011-12

The average year-on-year aggregate inflation (7.4%) and food inflation (7.2%) were at comparable levels during FY91 to FY06, but post 2007 financial crisis food inflation has always exceeded aggregate retail inflation by more than two percentage points, barring two years of FY11-12 in which crude oil prices rose sharply taking fuel component of total inflation with itself (see Fig. 1). The effect of decline in global food prices in FY10 was not transmitted into the domestic market as the domestic food prices grew by more than 15% in that year primarily due to excessive stock hoarding by FCI in the wake of the 2007-08 global food crisis. With a moderate gain in FY12, food prices started rising again as the household inflation expectations remained at persistently high levels during FY10-14, a period which witnessed a surge in crude oil prices and political upheavals with state and general elections being held in 2013-14. Various government interventions including significant hikes in MSP, pre-election policy announcements involving food and fertilizer subsidies and other populist measures caused ballooning of central and state fiscal deficits. These measures not only caused inflation in the immediate years but also prolonged the inflationary pressures in the economy resulting in double-digit food inflation during FY13-14. With the creation of an inflationary spiral of elevated inflation expectations[11] and food inflation levels transmitting into core inflation and wages[12],

[11] According to RBI (2014), a one percent increment in food inflation is followed by an immediate rise in one-year-ahead household inflation expectations by half percentage points, the effect of which persists for eight quarters.
[12] The influence of Indian food inflation on price expectations and wage setting create large second-round effects on core inflation (Anand et al. 2014)

the aggregate retail inflation remained at uncomfortably high levels during 2010-14 averaging more than ten percent. Within the food items, the highest inflation was observed among pulses (a major source of protein) with a three-fold rise in prices, whereas the price of overall food basket has doubled in the last ten years.

## 4.3 Results and Discussion

The empirical results are based on the methodology described in Section 3. To test stationarity of time series under consideration, ADF (Augmented Dickey Fuller) and KPSS (Kwiatkowski–Phillips–Schmidt–Shin) tests were conducted. Accordingly, the results of these tests, as reported in Table 2, indicate that all variables are level stationary.

**Table 2** Results from unit root tests

| Variables | ADF test statistic (level)<br>$H_o$: Variable follows a unit root process | KPSS test statistic (level)<br>$H_o$: Variable follows a stationary process |
|---|---|---|
| FCPI | -3.067** (0.029) | 0.129 |
| MonsDev | -5.003*** (0.000) | 0.305 |
| MSP | -2.884** (0.047) | 0.189 |
| FAO | -3.275** (0.016) | 0.184 |
| FD | -2.968** (0.038) | 0.106 |
| FWI | -1.995 (0.289) | 0.15 |
| AgriInput | -2.613* (0.090) | 0.421* |
| ProteinExp | -5.514*** (0.000) | 0.184 |

*** implies 1% significance level, ** implies 5% significance level, and * implies 10% significance level. KPSS test: Critical values are 0.347 (10%), 0.463 (5%), and 0.739 (1%). ADF test: Figures in parentheses are MacKinnon *p*-values.

Next, we test the hypotheses that each of the other variables does not Granger-cause the dependent variable in the VAR equation. Accordingly, Table 3 reports the Wald test statistic for all the other variables under the null hypothesis that the coefficients on all the lags of a regressor variable are jointly zero. The results show that all the variables included in the VAR model are statistically significant and are Granger-causing food inflation.

**Table 3** Results of Granger causality Wald tests (regressor causing FCPI in a Vector Autoregressive framework)

| Regressor | MSP | FWI | AgriInput | FAO | ProteinExp | FD | MonsDev |
|---|---|---|---|---|---|---|---|
| chi-squared test statistic | 15.613** (0.016) | 58.173*** (0.000) | 83.032*** (0.000) | 53.328*** (0.000) | 19.451*** (0.003) | 30.377*** (0.000) | 13.436** (0.037) |

$H_o$: Regressor does not Granger-cause FCPI. Figures in parentheses are p-values.

*** implies 1% significance level, ** implies 5% significance level, and * implies 10% significance level.

For *step 3*, the BRT model has been constructed over 50,000 trees with a learn rate of 0.0001 and subsample fraction of 0.95. To capture multi-variable interactions in the given small dataset, six maximum nodes per tree are allowed and minimum number of observations in terminal nodes is kept at three. The developed model fits[13] the actual data quite well (see Fig. 2) with a R-squared value of 99.1 percent and MSE (mean squared error) flatlining beyond 30,000 trees (Fig. 3). Readers should refrain from interpreting these error measures as *'predictive accuracy/performance'* measures traditionally reported in predictive analysis, and instead view these as parameters which are required to cross a certain value or stabilize for this hybrid approach to work reliably and, certainly not to be confused with hyperparameters of machine learning. An encouraging instance depicting unconventional usage of model *fit* is illustrated in Athey and Imbens (2015), p. 3, wherein the authors propose a novel method based on machine learning of regression trees to measure robustness of parameter estimates; one of the steps involved there required selection of split point offering greatest improvement in model fit.

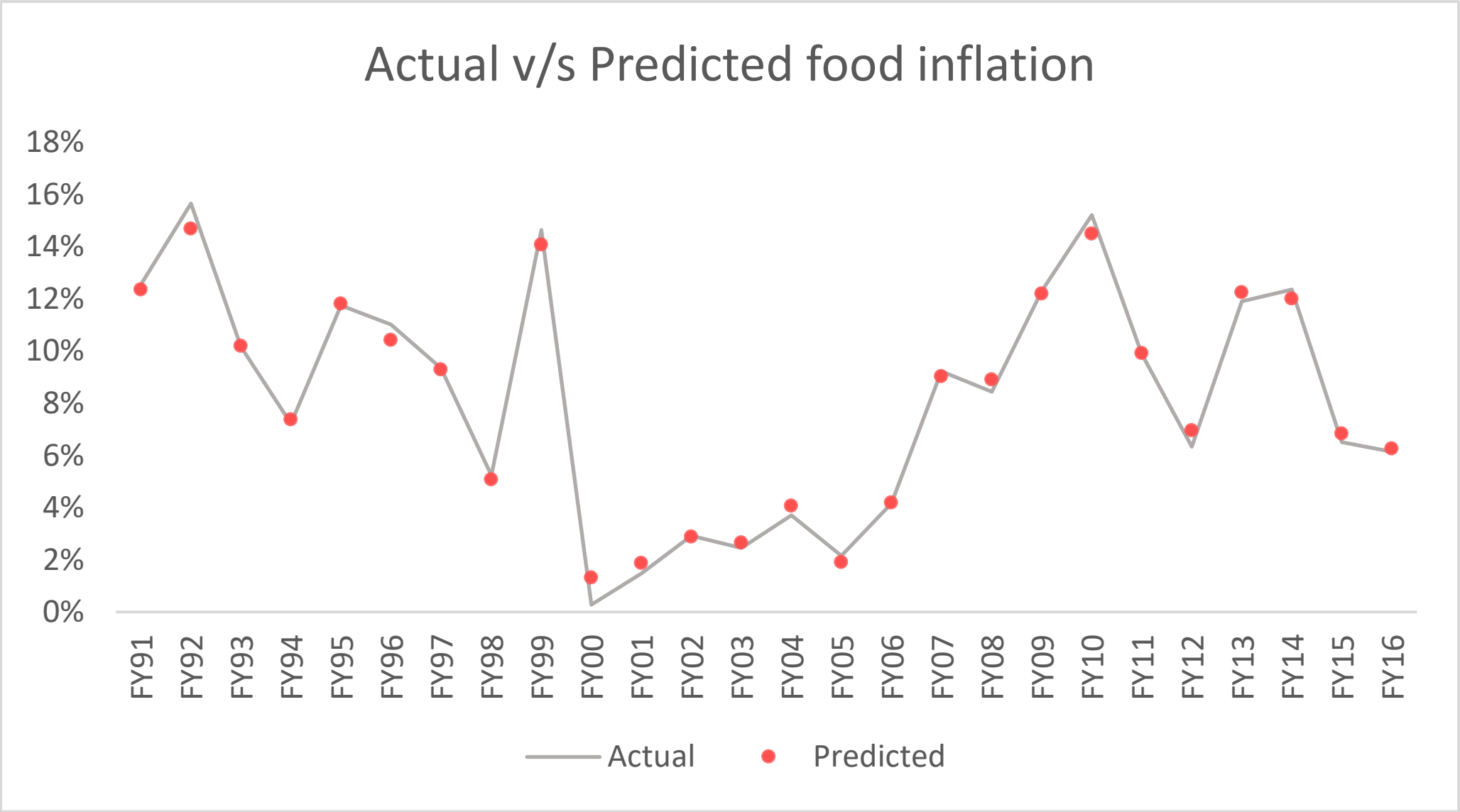

**Figure 2.** Comparison of output of BRT model to actual data

[13] Readers should not confuse this with *'overfitting'* as the model in question here is exploratory (no independent testing) in nature, not a predictive one.

The results (see Fig. 4) indicate that all variables have non-zero importance[14] in explaining the inflation in food prices and contributions are fairly distributed among the variables, except probably FAO, with MSP contributing the most to the performance of BRT, followed by farm wages/Protein expenditure. Note that splitting and permutation-based importance measures both show similar kind of scores and ranking, with position reversals taking place for farm wages/protein expenditure and monsoon deviation/fiscal deficit.

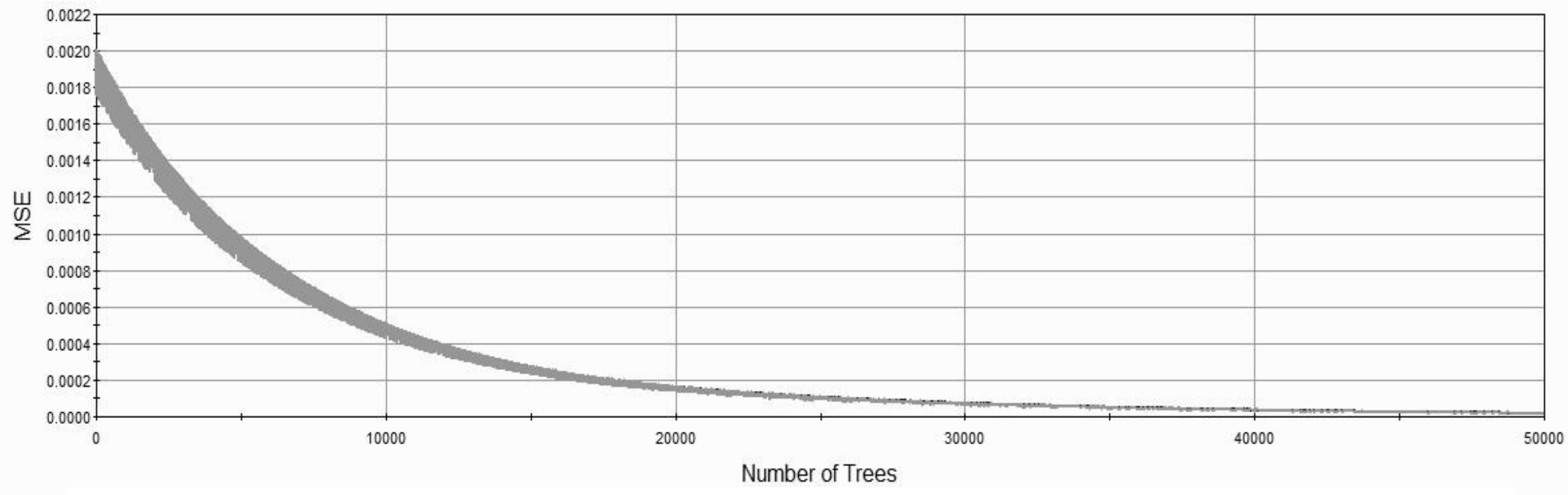

**Figure 3.** Change in MSE with growing trees

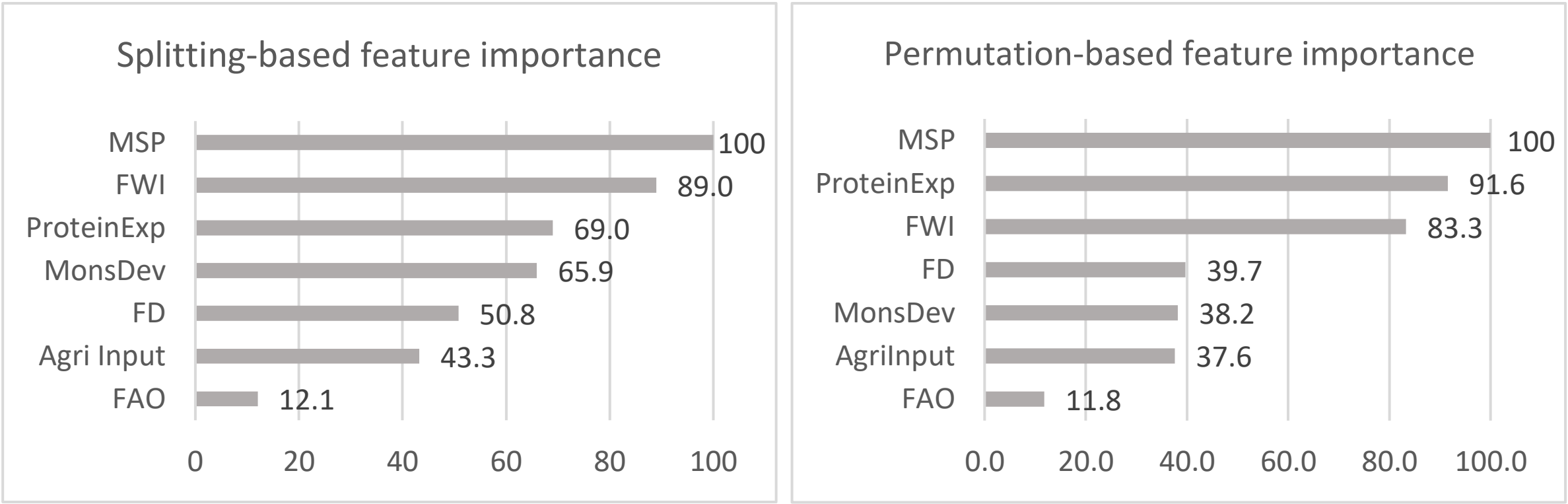

**(a)** **(b)**

**Figure 4.** Relative importance of explanatory variables

Minimum support prices carry most importance as they set the floor for market prices, but during years with substantial hikes, it eventually ends up setting market prices directly (Rajan 2014; Mishra and Roy 2011). The role of MSPs in guiding food inflation is fairly large as the crops covered under MSP constitute more than a third of all-India food consumption basket[15].

---

[14] Note that, *'importance'* here is not same as *'statistical significance'*.

[15] *Source*: All India Weights of different Sub-groups within Consumer Food Price Index - 2012 series

Additionally, hikes in MSP not only elevates the inflation expectations but combined with PDS and employment guarantee schemes they set indirect inflationary pressures in the economy as well, by burdening the exchequer with bloated subsidy bills and inducing a wage-price spiral. The $2^{nd}/3^{rd}$ most important variable, farm wages, have recorded a sharp rise especially from FY08 onwards and during this period the annual food inflation has averaged above 9 percent. A rise in farm wages inflates the cost of production and is believed to cause a wage spiral in the economy by raising the benchmark *'reservation wage'* — the lowest rate that workers are prepared to accept for jobs across sectors which could possibly increase demand for food as well (Gulati and Saini 2013). Change in farm wages disturbs the food price equilibrium in an unbalanced way as low rural wages suppress demand only to a certain extent owing to the relatively low income elasticity for food expenditure of rural households (which constitutes 70 percent of Indian population). On the other hand, a rise in rural wages not only boosts demand but also rejigs the food basket towards higher value and protein-rich items, thereby disturbing both the demand (strong rural demand) and supply (high labor costs) side of price equilibrium. With rising incomes, dietary preference has shifted in accordance with Bennet's law towards more nutritious and high-value food items away from starchy cereals. The share of protein-rich food items, both plant and animal, in total food consumption has risen to almost 40 percent (Rajan, 2014). The prices of nutrition-rich items, including vegetables and fruits, have recorded a higher inflation than cereals owing to suppressed growth in the supply of these items, causing a mismatch with rapidly growing demand (Mohanty 2011 and 2014). Evidence from the literature (Rajan 2014; Gokarn 2010 and 2011) asserts that protein and high-value food items have become prominent determinants of overall food inflation in the recent times.

The very nature of Indian economy puts in perspective, the significance of rains in making a good crop year, especially southwest monsoon lasting from July to September (Mitra and Chattopadhyay, 2017). More than half[16] of India's net sown area, land which is cropped at least once a year, still remains unirrigated and relies on water that rains down from the clouds. A deficient monsoon builds inflation expectations and affects the agricultural output and hence food prices (Mohanty 2010; 2014). In a developing economy like India, where demand for food is relatively inelastic and short-run price elasticities of agricultural supply are still lower, any shift in demand arising out of macroeconomic policy change, such as fiscal or monetary stimulus, could distort food prices significantly. A similar situation arose post 2007-08 financial crisis, prior to which India had successfully brought down its Fiscal Deficit from almost 10 percent of GDP in FY02 to 4 percent in FY08 following the impressive fiscal consolidation with the passage of 2003 FRBM Act. As the fears of severe unemployment and recession grew in the wake of the 2007-08 financial crisis, G7+5 countries, as well as IMF, favored a fiscal stimulus up to 2 percent of GDP as a pathway out of the feared recession. There was a striking difference between the expansionary fiscal policies adopted in India and other countries, for instance in China, nearly 40 percent of the fiscal stimulus was directed toward investment in infrastructure development whereas, the Indian fiscal package was more focussed on stimulating demand by granting direct subsidies (ranging from food, fertiliser, energy subsidies to even flat waivers in agricultural debt[17]), expansionary income support schemes like MGNERGS (Mahatma Gandhi National Rural Employment Guarantee Scheme) and generous pay hikes announced in Sixth Pay Commission for state servants. These welfare and employment guarantee schemes imparted substantial amounts of liquidity and purchasing power, particularly to rural households, boosting demand for food items (Rakshit 2011) but with several supply bottlenecks in place, particularly, stagnant productivity and subpar infrastructure, the situation soon got transformed into demand-pull inflation.

[16] *Source:* Ministry of Agriculture & Farmers Welfare, Government of India.

The cumulative non-labour agri-input price index has recorded more than 80 percent rise during FY05-FY16 which has emerged as a fairly important variable that is supposed to indirectly affect market prices through MSPs as, apart from labor costs, CACP (Commission

[17] Just before 2009 general elections, then ruling United Progressive Alliance (UPA) government waived the repayment of loans made to small and mid-sized farmers which, some commentators believe, helped the coalition get re-elected (Rajan 2011)

for Agricultural Costs and Prices) takes other factors of agricultural production into account, as well, while recommending MSPs to the government. This possibly explains the primacy of MSP among all variables. The relatively low importance of FAO index may in part relate to agricultural trade policies adopted by India which prevent transmission of global food price short-term volatility into the domestic markets and allow only long-term trend in global food prices to be captured into the recommendations made by CACP for setting MSPs.

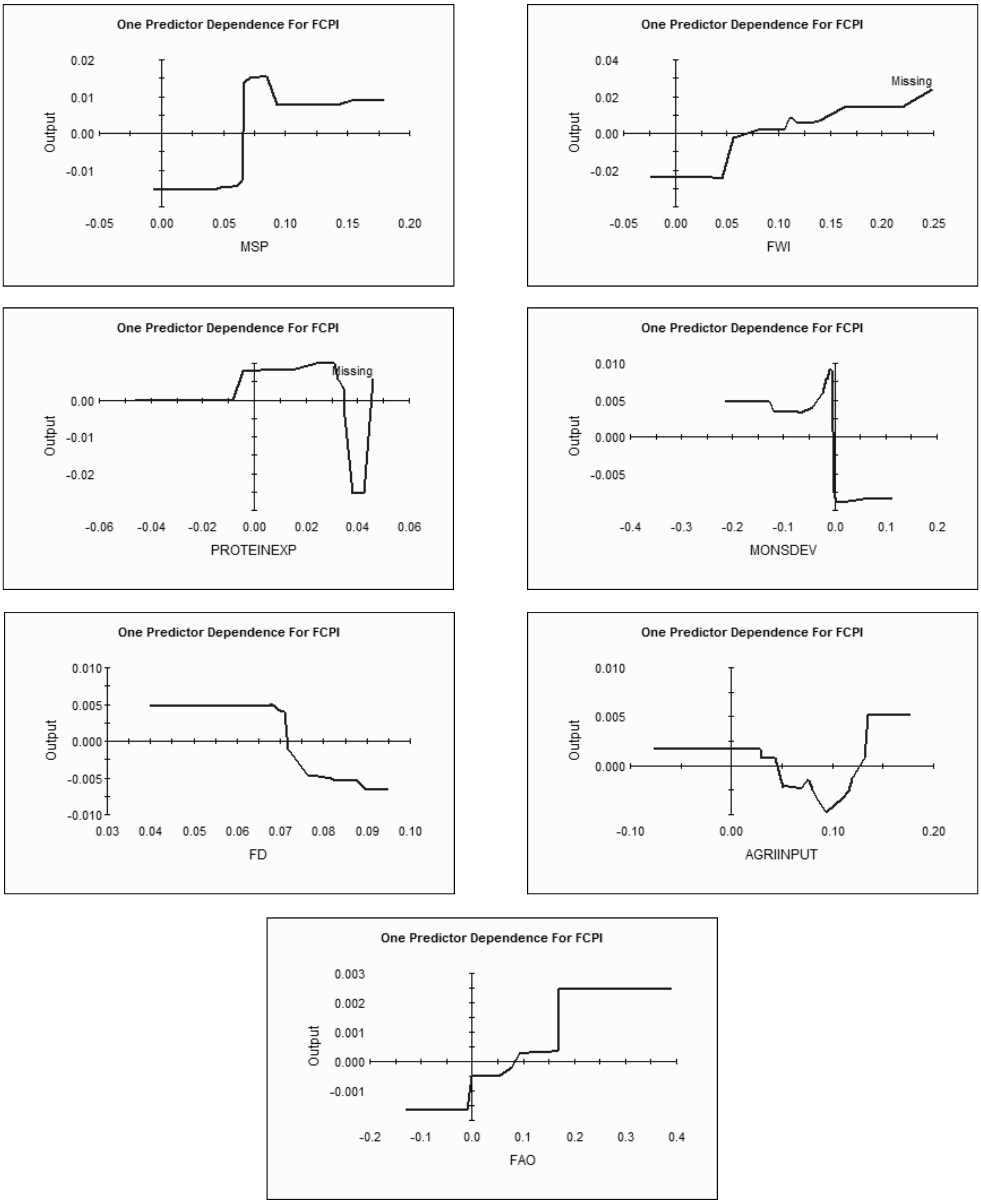

**Figure 5.** Univariate partial dependence plots for all predictors

Additionally, BRT model allows visualisation of the relationships between a single independent variable and the response variable via univariate partial dependence plots (Friedman, 2001; Friedman and Meulman, 2003) wherein, the independent variable of interest is varied over its range while the remaining variables are fixed at joint set of values sampled from the dataset to produce an instance of the response dependence curve (not be confused with causality). This process is replicated for all learn records, sampling new joint set of fixed values each time, creating a family of partial dependence curves which are then averaged and centered so as to generate the final partial dependence plot (PDP). The comparable scales of vertical axes in all plots (Fig. 5) indicate that all the independent variables are important, including FAO, and the hierarchy is not too intense which is corroborated in Fig. 4 as well.

## 4.4 Robustness checks

In order to test the robustness of the proposed hybrid approach, we focus on multicollinearity. The correlation matrix, as shown in Table 4, indicate that there does exist multicollinearity in our illustrative example. As discussed in Section 2, relative importance of the independent variables having the weakest correlation with dependent variable may get subdued in the presence of multicollinearity. As reported in Table 4, ProteinExp, AgriInput and MonsDev are the independent variables that have weak correlations with FCPI (the dependent variable). However, ProteinExp is the 3rd/2nd most important explanatory variable and possess high relative importance scores (see Fig. 4). The relative importance scores of the other two explanatory variables, AgriInput and MonsDev, also indicate towards an absence of a downward bias in the computed relative importance scores of independent variables that are weakly correlated with the dependent variable.

As highlighted in Section 2, a major shortcoming of the *relative weights analysis* approach (Johnson, 2000) was: when multicollinear variables have similar correlations with dependent variable also, relative weights could get artificially inflated. Here we observe that MSP and FAO are highly correlated and have similar levels of correlation with FCPI (see Table 4), but still, FAO emerges as the lowest ranked variable (see Fig. 4) and has a significantly lower importance score than other variables. Similarly, FD and MonsDev are correlated and have similar levels of correlation with FCPI, but both the independent variables have moderate relative importance scores. Thus, we do not observe inflation of relative importance scores (as obtained from the hybrid approach) for multicollinear independent variables with similar levels of correlation with dependent variable. The above results show that the proposed hybrid approach is superior to the conventional econometric methods of ascribing relative importance, at least in terms of handling multicollinearity.

**Table 4** Correlation matrix

| | FCPI | MSP | FWI | AgriInput | FAO | ProteinExp | FD | MonsDev |
|---|---|---|---|---|---|---|---|---|
| FCPI | 1 | | | | | | | |
| MSP | 0.555 | 1 | | | | | | |
| FWI | 0.628 | 0.450 | 1 | | | | | |
| AgriInput | -0.074 | -0.066 | -0.057 | 1 | | | | |
| FAO | 0.412 | 0.589 | 0.193 | 0.336 | 1 | | | |
| ProteinExp | 0.068 | 0.250 | 0.109 | -0.128 | 0.073 | 1 | | |
| FD | -0.263 | -0.428 | -0.207 | -0.003 | -0.542 | -0.030 | 1 | |
| MonsDev | 0.195 | 0.315 | 0.166 | 0.107 | 0.379 | 0.021 | -0.564 | 1 |

## 4.5 Takeaways for Policy

Policy practitioners often grapple with assigning priorities to the underlying variables and in the lack of a quantitative approach end up making decisions based on political ideologies or some other subjectivity. In this regard, the proposed hybrid approach which ascribes relative importance to underlying variables could certainly prove helpful in structuring policy priorities. In the present study, MSP emerged as the most important factor affecting inflation in food prices and should be dealt accordingly. The efficacy of MSP hikes to act as production incentive becomes questionable in itself, in the presence of inflationary effects of cost-indexed MSP on farm input costs and subsidies on the same farm inputs. The use of incentive schemes like MSP in boosting agricultural production may be limited as Rajan (2014) argues that “the gains from MSP increases have not accrued to the farm sector in full measure on account of rising costs of inputs”. There is a need to evaluate whether an exclusive policy — providing price support for output or subsiding inputs — would be a sufficient stimulus for agricultural production. The principal role of MSPs should be the alignment of domestic prices along the long-term trend in international prices.

The 2nd/3rd important factor, rising farm wages, is a positive sign and going forward, the focus should certainly not be on stalling this rise but rather on bringing down the pay-productivity gap. A high agricultural labor productivity is essential for supplementing the higher demand arising out of increase in wages. The desired productivity move is achievable through adoption of advanced agronomic technologies, investment in mechanization, and extension of farms. However, the growing dependence on productivity for raising production to meet domestic demand might not help in easing inflationary pressures in the short to medium term and may cause prices to rise even more as the introduction of new technologies increases the average cost of production during adoption years. This might lead to a situation where inflation in food prices would sustain in a period of rising agricultural output.

The next factor in relative importance, shift in the dietary pattern towards pulses and other protein-rich items, is certainly a welcoming sign, but to avoid a *'calorie catastrophe'*, the Indian government needs to offer remunerative procurement prices for pulses, which would not only incentivise farmers with small un-irrigated plots but would also encourage cultivators with access to capital and irrigation to invest in pulses. The recently launched national mission (PMKSY) to expand the cultivable area under irrigation is a welcome step forward in the regard of reducing monsoon dependency of Indian farms, the 4th/5th important variable emerging out of the present study. In regard to the next important variable, a renewed Indian government's commitment to fiscal consolidation is commendable and would certainly support the disinflation process going forward. The ability of Indian government to bring in major agricultural policy reforms would not only determine the trajectory of food inflation in the coming years but also have a lasting effect on agricultural growth and ultimately, rural poverty in India.

# 5. Conclusions

Even when the primary aim of any study is explanation or establishing causal inferences, researchers should consider conducting relative importance analyses of their explanatory variables, as a valuable supplement to their primary analyses. In this article, I proposed a hybrid approach based on machine learning algorithms and conventional econometrics as an addition to an unsettled, to date, account of methods to assess the relative importance of independent variables. The results obtained from the application of this methodology to determinants of food inflation in India proved to be coherent with economic understanding of food policy in India and efficacious in making relevant policy recommendations.

As we proceed to the end of this article, I would like to highlight a few caveats, some of which carry significant implications for future avenues of research. Firstly, the aim of this article is not to establish the primacy of machine learning algorithms among other approaches but to suggest an alternative in an otherwise scant literature of relative importance. A comparison between the proposed hybrid approach, dominance analysis and relative weights using datasets from eclectic sources (similar to exercise undertaken by Johnson, 2000) could be an interesting future research avenue. Econometricians, and statisticians in general, ought to explore more and newer approaches for relative importance, among which a supremacy can't be established, theoretically. A tenable way to settle the debate on supremacy among these approaches would be to conduct a series of Randomized Controlled Trials (RCT) with one control and '*n*' treatment groups, where '*n*' is the number of independent variables (standardized) acting as sole treatments in their respective treatment groups, then take the magnitude of average treatment effects (again standardized) as a measure of relative importance of these '*n*' independent variables. The relative importance approach which imitates the results from RCT best could be proclaimed as supreme. This proposal might serve as an important direction for future research.

The hybrid approach proposed here should be viewed as a supplementary tool to expand the interpretation of explanatory results obtained from econometric methods and not as a replacement of those methods itself. The hybrid approach is not a substitute for what conventional econometrics already does well, for example, it is not appropriate for tasks like identifying the best subset of variables for dimensionality reduction or identifying suppressor variables and interaction effects. However, when a ML algorithm is used for relative importance assessment, it becomes necessary to ensure that the value and interpretation of the computed importance scores truly reflect the importance of variable and are not affected by some idiosyncrasy or other present in the dataset. Strobl et al. (2007) found that both the measures of variable importance (splitting and permutation) are not reliable when variables vary in their scale of measurement for original random forest algorithm (a type of tree-based ensemble method); hence, it's always a good practice to normalize or transform the variables to a scale-independent form when dealing with relative importance problem, as illustrated in Section 4 of this article. When applying this hybrid approach, it might still be necessary to consider the correlations among variables for some of the algorithms as Strobl et al. (2008) argue that the variable importance measures show a bias towards correlated variables in the case of random forests. However, in the case of BRT it appears that correlations might not be causing much of a problem as, in the food inflation example of Section 4, MSP and FAO were the most correlated variables ($r = 0.6$), but still, FAO emerged as the lowest ranked variable and had a significantly lower importance score than other variables. This is not a wholly tenable argument and requires further study on the reliability of importance scores derived from different ML algorithms in the presence of multicollinearity.

Regarding interpretability of results of this hybrid approach, users should remain cautious while judging if '*small*' differences between relative importance scores (permutation or splitting) are meaningful since there are no built-in protections like *p-value*, associated with this approach. This is certainly a deficiency plaguing the conflation of ML and econometrics in the present times, as rightly pointed out by Athey (2019). Construction of valid confidence intervals or alternate metrices to ensure fairness and comparability for such ML-based approaches could be an exciting line of future research[18].

As a closing remark, I would reiterate that prediction is not same as explanation or causal inference (Shmueli, 2010) and the type of uncertainties[19] associated with them are different (Helmer and Rescher, 1959) and owing to these, the usage of machine learning algorithms in econometrics, which are otherwise, used in predictive analytics, might seem unreasonable to some readers but it's important to bear in mind that the distinction of ML as prediction vs causal inference tool is not defined, per se, and there is an emerging body of literature advocating the employment of ML in causal inference, viz., Grimmer (2015), Varian (2016) and Mullainathan and Spiess (2017). And in regard to the problem of relative importance, Johnson and LeBreton (2004) rightly points out that there is no unique mathematical solution to the problem; the existing and any new approach must be evaluated on the basis of the logic and sensibility behind their development, and whatever limitations/incoherencies can be identified in their construct or results produced by them. However, what makes conceptual sense is highly subjective. In view of that, the proposed hybrid approach, which blends ML and conventional econometrics to assess relative importance of variables, is no more (or less) arbitrary than widely recommended (Johnson and LeBreton, 2004; Stadler et al., 2017) approach of taking orthogonal transformation on variables.

# Supplementary Material

The dataset for replication of results presented in this study could be accessed at

http://dx.doi.org/10.7910/DVN/JUNBRQ.

[18] A notable contribution in this direction has been made by Wager & Athey (2018), wherein the authors propose a method for constructing asymptotic confidence intervals in the context of *'causal forests'*—a modification of random forests for causal inference.

[19] Predictive exercises pay more attention to model uncertainties as opposed to explanation which considerably focusses more on statistical uncertainty and standard errors (Einav and Levin, 2014)

# Appendix A

## Mathematical Formulation of BRT

Gradient tree boosting employs decision trees of constant size as basis functions, $h_m(x)$, referred to as weak learners in this context, for generating additive models as the following:

$$F(x) = \sum_{m=1}^{M} \gamma_m h_m(x)$$

wherein $M$ and $\gamma_m$ represent the total number of trees employed and step length respectively. The model is constructed in a progressive stepwise manner,

$$F_m(x) = F_{m-1}(x) + \gamma_m h_m(x)$$

At each step, the next decision tree $h_m(x)$ is chosen to minimise the given loss function, $L$, for the current model $F_{m-1}(x)$ and its fit $F_{m-1}(x_i)$:

$$F_m(x) = F_{m-1}(x) + \arg\min_h \sum_{i=1}^{n} L(y_i, F_{m-1}(x_i) - h(x))$$

The initial model $F_0$ is specified by the type of loss function used. The idea behind gradient boosting is to solve this minimization problem numerically through steepest descent. The steepest descent is identified as the negative gradient of the given loss function calculated at the current model $F_{m-1}$,

$$F_m(x) = F_{m-1}(x) + \gamma_m \sum_{i=1}^{n} \nabla_F L\big(y_i, F_{m-1}(x_i)\big)$$

Now, the step length $\gamma_m$ is selected by means of line search:

$$\gamma_m = \arg\min_\gamma \sum_{i=1}^{n} L\left(y_i, F_{m-1}(x_i) - \gamma \frac{\partial L\big(y_i, F_{m-1}(x_i)\big)}{\partial F_{m-1}(x_i)}\right)$$

In addition to conventional BRTs, a simple regularization strategy, as proposed by Friedman (2001), has been employed to improve the accuracy of the model which scales the contribution of each tree by a factor $\vartheta$, also known as learning rate:

$$F_m(x) = F_{m-1}(x) + \vartheta\gamma_m h_m(x)$$

Stochastic gradient boosting, which combines gradient boosting with bagging, further improves the performance of model; wherein at each iteration, a subsample of data is randomly drawn (without replacement) from the available dataset (Friedman, 2002).

## Model parametrization

The calibration of developed BRT model is done through the following parameters,

i. *Number of trees:* The number of trees/iterations determines the model complexity. A large number of trees is recommended to exhaust the internal data structure and bring the mean squared error to statistically acceptable levels.

ii. *Learn rate / Shrinkage:* Slow learning rate improves predictive performance at the expense of increased computation time, however, smaller shrinkage values are recommended while growing a large number of trees.

iii. *Maximum nodes/splits per tree:* The number of splits determines the tree complexity and order of interactions between predictor variables. In general, a tree with *k* nodes can capture interactions of order *k*. Hastie et al. (2009) recommends four to eight splits for most cases.

*iv.* *Minimum number of observations in terminal nodes:* This parameter should be kept below five for small datasets

*v.* *Subsample fraction:* The rate of subsampling decides the proportion of learn sample to be used at each modeling iteration. Subsampling fraction close to 1 makes the model computationally intensive, however, values between 0.9 to 0.95 are recommended when dealing with small datasets.

*vi.* *Regression Loss criterion:* Among all kinds of loss functions, the natural choice for regression is least squares owing to its overlying computational attributes. For least squares, the initial model $F_0$ is specified by the mean of the target values.